\documentclass{amsart}
\usepackage[T1]{fontenc}
\usepackage{lmodern}
\usepackage{upquote}
\usepackage{amssymb}
\usepackage{natbib}
\usepackage{amsbsy}
\usepackage{listings}
\usepackage{amsaddr}
\usepackage{amsfonts}
\usepackage{amsmath}
\usepackage{url}
\begin{document}

\title[Using RngStreams with C++ and R] {Using RngStreams for Parallel Random Number Generation in C++ and R}

\author[Karl, Eubank, Milovanovic, Reiser, and Young]{Andrew T. Karl}
\address{Adsurgo LLC, Denver, CO}
\author{   Randy Eubank \and
        Jelena Milovanovic \and
         Mark Reiser \and
        Dennis Young
}
\address{Arizona State University, Tempe, AZ}

\begin{abstract}
The \texttt{RngStreams} software package provides one viable solution
to the problem of creating separate random number streams for
simulations in parallel processing environments. Techniques are
presented for effectively using \texttt{RngStreams} with \texttt{C++}
programs that are parallelized via \texttt{OpenMP} or \texttt{MPI}.
Ways to access the backbone generator from \texttt{RngStreams} in
\texttt{R} through the \texttt{parallel} and \texttt{rstream} packages
are also described.  The ideas in the paper are illustrated with both a
simple running example and a Monte Carlo integration application.
\end{abstract}
\maketitle
\section*{NOTICE}
The original article appeared in \textit{Computational Statistics}
(2014), 29:1301--1320, \url{https://doi.org/10.1007/s00180-014-0492-3}.
This revised preprint corrects notation and bibliographic details,
clarifies stream allocation, the \texttt{mclapply} example and importance
sampling, and identifies the application data as LSAT Section VII.
The numerical examples and timing results
are unchanged and retain the software context of the original study.

The first author used Codex (GPT-6 Astra) to audit this paper and his
other papers on arXiv for typographical and mathematical errors and to
refine the exposition. No errors affecting the reported results or
conclusions of this paper were identified. GPT-6 Astra suggested
corrections, including prose revisions for clarity, which the first
author reviewed and accepted; these changes are incorporated in this
revised preprint.

\section{Introduction}
\label{intro}
Simulation studies are in many respects the ideal application for
parallel processing. They are ``naturally parallel'' in the sense
that computations may generally be conducted without the need for
inter-processor communication or modification of the base algorithm.
Thus, near linear speedup can be realized in many cases that occur in
practice. However, in order for a division of labor across multiple
processors to be productive, the random number streams being used by
each processor must behave as if they are independent in a
probabilistic sense. Methods now exist that, when used properly, allocate non-overlapping streams with theoretical and statistical support for treating their outputs as independent in simulation. We explore one
such approach in this paper that is available through the
\texttt{RngStreams} software package.

The effect of stream dependence is easy to demonstrate as seen in the
output from an \texttt{R} session exhibited below.
\lstset{language=R, basicstyle={\ttfamily\footnotesize}}
\begin{lstlisting}
> x <- NULL
> for (i in 1:200) {
+   set.seed(123)
+   .Random.seed[3:626] <- as.integer(.Random.seed[3:626] +
+                                     (i - 1) * 100)
+   zTemp <- rnorm(1000)
+   x <- c(x, sqrt(1000) * mean(zTemp))
+ }
> set.seed(123)
> y <- NULL
> for (i in 1:200) {
+   zTemp <- rnorm(1000)
+   y <- c(y, sqrt(1000) * mean(zTemp))
+ }
> cat(shapiro.test(x)$p.value, shapiro.test(y)$p.value, "\n")
0.0001438936 0.4466189
\end{lstlisting}
Following along the lines of \citet{hech}, we have generated sets of
1000 random numbers from the standard normal distribution using 200
different random number streams deriving from \texttt{R}'s default
Mersenne Twister (MT) random uniform generator. The streams are
constructed by using seeds (vectors of length 624) that are spaced 100
units apart in each component. The mean from each stream is then
rescaled to produce numbers that have ostensibly been sampled from a
standard normal distribution. However, the Shapiro-Wilk test firmly
rejects the normality hypothesis. In contrast, data from a similar
process that permits the normal evolution of the MT states easily pass
the Shapiro-Wilk evaluation.  Although the relationship between
streams will likely be much more complex in situations that arise in
practice, the message remains the same: inter-stream dependence can
sabotage a parallel random number generation scheme.

\citet{hill} reviews available methods for parallel random number
generation, including random spacing (workers are initialized to
randomly spaced positions on the period of the same generator by
assigning different, randomly generated seeds), sequence splitting
(dividing a sequence into non overlapping contiguous blocks), cycle
division (a more involved technique for sequence splitting where the
period of a generator is deterministically divided into segments) and
parametrization (parameters associated with a generator are varied to
produce different streams). The random spacing method carries a risk
of overlap between random streams on different processors with an
unlucky sampling of seeds, although the risk is small when using
generators with large periods.

MT provides one example of a long-period generator with the period
$2^{19937}-1$ \citep{MT}.  This feature along with its widespread
availability has made MT a popular choice for parallel generation via
random spacing. In spite of this, there are reasons that make the
development and use of other parallel generation methods
worthwhile. For example, it is generally advisable to repeat
simulations with different generators to ensure that the results are
not an artifact of the structure of a particular generator
\citep{gentle}. So, there is value in having other available generator
options for this purpose. In addition, a method such as cycle division
comes with a guarantee that the streams will not collide, whereas
random spacing only renders such events unlikely. \citet{lemieux}, for
example, advocates against random spacing and recommends using
generators that are guaranteed not to overlap. As a case in point, she
directs us to the work of \citet{failure} who show that many modern
random number generators exhibit correlations if their initial states
are chosen using another linear generator with a similar modulus.

Issues that arise when using linear congruential generators for
parallel random number generation have been well documented. Random
spacing and sequence splitting may result in random number streams
that exhibit undesirable dependence properties (e.g.,
\citet{matteis1}, \citet{matteis2}, \citet{anderson}, and
\citet{etacher}). Similar results have been quantified for some
nonlinear generators. For example, \citet{matteis3} show that long
range correlation is present in random number streams produced by
sequence splitting with any generator of the form $x_n = f(x_{n-1})
\bmod 2^m$ if the generator is such that the period halves when the
modulus halves.

Many of the problems with congruential generators can be avoided by
using combined multiple recursive generators (CMRG) with cycle
division. A CMRG is roughly equivalent to a single MRG with a period that can be as large as the product
of the periods of the generators used in the combination
\citep{crganal}. When combined appropriately (e.g., as in
(\ref{eq:combform}) below), the lattice structure inherent to multiple
recursive generators is effectively blurred \citep{crglattice}. With a
good choice for the parameters (e.g., \citet{crgparameters}), these
combined generators have also fared well when subjected to statistical
tests as in \citet{testart}. As a case in point, the MRG32k3a
generator that provides the backbone generator for the
\texttt{RngStreams} package developed by \citet{rngstream} is known to
have good theoretical and statistical properties. It is cited by
\citet{lemieux}, along with MT, as a generator that can be ``safely
used''.

Of course, cycle division becomes an option only if it can be done
efficiently. For multiple recursive generators \citet{lecuyer} has
shown that there are computable matrices that can be used to advance
the state of the generator to any specified point in its associated
random number stream. This feature is what makes the combination of
generators easy to use in a cycle division paradigm. The
\texttt{RngStreams} package gives an implementation of this approach
in the context of its MRG32k3a generator. \citet{mtjump} have recently
developed a jump ahead method for MT. The number generation step is
two to three times faster for each stream than with
\texttt{RngStreams}; but, the sequence splitting itself is roughly
1000 times slower.

Another noteworthy package that provides functionality for parallel
random number generation is the \texttt{SPRNG} package of
\citet{sprng}. In contrast to \texttt{RngStreams}, it produces
different random number sequences for the processes via
parametrization of a single type of generator. For example, one of the
package's featured generators is a multiplicative lagged Fibonacci
generator whose states fall into $2^{1007}$ different equivalence
classes of generators (each of period $2^{81}$) that provide the
different streams.

In this paper we focus on the use of the \texttt{RngStreams}
package. There are several reasons for this. For example,
\citet{rngstream} in reference to the \texttt{SPRNG} package state
that it is ``not supported by the same theoretical analysis for the
quality and independence of the different streams'' as
\texttt{RngStreams}. There are also practical difficulties that arise
in using \texttt{SPRNG} due to its ties to \texttt{MPI} and a rather
complex implementation that necessitates the creation and linking of a
compiled library. In contrast, the \texttt{RngStreams} package is
quite compact with all its source code available from
\url{http://www.iro.umontreal.ca/~lecuyer/myftp/streams00/}.

The \texttt{C++} implementation of \texttt{RngStreams} provides an
object oriented framework where objects are created which have an
associated method \texttt{RandU01} that produces uniform random
deviates. The overall result is that \texttt{RngStreams} is easy to
incorporate into \texttt{C/C++} programs using \texttt{OpenMP} or
\texttt{MPI} through simple include directives. This property extends
its applications from cluster down to desktop environments.

General treatments of parallel computing in \texttt{R} are provided by
\citet{schmid} and \citet{eugster}. In contrast to these papers, the
\texttt{R} segments of this article are directed toward the specific
task of parallel random number generation in the language. In this
regard, the \texttt{RngStreams} generators become available in
\texttt{R} through the \texttt{rstream} package \citep{rstream}. They
are also available for parallel stream allocation through the
\texttt{parallel} package, which is distributed with \texttt{R}.

The goal of this article is to illustrate how easily
\texttt{RngStreams} can be employed in parallel settings in
\texttt{C++} with \texttt{OpenMP} and \texttt{MPI} and in \texttt{R}
\citep{R} through the \texttt{rstream} and \texttt{parallel}
packages. Our initial presentation involves only the most elementary
features of these APIs. However, it is our belief that these simple
programs contain fundamental techniques that can provide a foundation
for safe parallel random number generation in much more complex code
that might evolve in practice. An application of Monte Carlo
integration in the context of an item trait model is used to support
this contention.

In the next section we describe the basic features of the
\texttt{RngStreams} generator. Then, in subsequent sections we address
its use in \texttt{C++} and \texttt{R}. These sections are connected
through a common example that allows us to illustrate both the
differences and similarities in using \texttt{RngStreams} under the
different APIs.

\section{RngStreams}
\label{sec;rngstreams}
The ``backbone'' generator for \texttt{RngStreams}, often referred to
as MRG32k3a, is a combination of the two multiple recursive generators
that produce the states
\begin{eqnarray}
\label{eq:rng1}
x_{1,n}&=&(1403580\times x_{1,n-2}-810728\times x_{1,n-3})\bmod (4294967087),\\
\label{eq:rng2}
x_{2,n}&=&(527612\times x_{2,n-1}-1370589\times x_{2,n-3})\bmod (4294944443)
\end{eqnarray}
at the $n$th step of the recursion given initial seeds
$\tilde{x}_{i,0} = (x_{i,-2},x_{i,-1},x_{i,0})^T, i = 1,2.$ The two
states are combined to produce the uniform random deviate $u_n$ via
the rule
\begin{eqnarray}
\label{eq:combform}
z_n&=&(x_{1,n}-x_{2,n}) \mod 4294967087,\\
u_n&=&\left\{\begin{array}{l l}
z_n/4294967088,&\text{ if } z_n>0, \\
4294967087/4294967088,&\text{ if } z_n=0.
\end{array}
\right.
\nonumber
\end{eqnarray}
This particular generator is capable of producing a random number
stream of period length (roughly) $2^{191}$ under the condition that
$x_{1,0},x_{1,-1},x_{1,-2}$ are not all 0 and are all less than
4294967087 and $x_{2,0},x_{2,-1},x_{2,-2}$ are all less than
4294944443 and not all 0. For use in a parallel context this long
cycle provides approximately $2^{64}$ non-overlapping streams of
length $2^{127}$. Avoiding overlap requires that each computation
remain within its allocated stream and that the allocation not wrap
around the generator's period.

The key to accessing different streams produced by
(\ref{eq:rng1})-(\ref{eq:rng2}) is the technique described in
\citet{lecuyer} that allows movement between streams through linear
transformations of the initial seeds using known
matrices. Specifically, if $\tilde{x}_{1,0}, \tilde{x}_{2,0}$ are the
initial seeds/states for the two generators, at the $n$th step of the
recursion the state for generator $i$ is $\tilde{x}_{i,n} = (A_i^n
\bmod m_i)\tilde{x}_{i,0}\bmod m_i, i = 1, 2,$ for known $3 \times 3$
matrices $A_1, A_2$ with $m_1 = 4294967087, m_2 = 4294944443$.  The
powers of the matrices $A_1, A_2$ can be computed explicitly and, in
particular the matrices $A_1^{2^{127}}, A_2^{2^{127}}$ that are needed for
movement between the different streams of the generator have been
calculated and are contained in an anonymous namespace in the
\texttt{RngStream.cpp} file. This feature is used in the \texttt{MPI}
section of our paper.

The initial state of the package is set using the function
\texttt{SetPackageSeed} that has prototype
\lstset{language=C++,basicstyle={\ttfamily\footnotesize},showstringspaces=false}
\begin{lstlisting}
  static bool SetPackageSeed (const unsigned long seed[6])
\end{lstlisting}
with \texttt{seed} containing the six initial seeds for the generator
that must be supplied by the user. The first \texttt{RngStream} object
that is created will use this initial seed. Subsequent objects will
then be started on the next stream; i.e., they will have initial
``seeds'' that have been advanced $2^{127}$ states from the initial state of
the previous object's generator. Our aim in the sequel is to
perform some simple experiments that will illustrate how the seeds are
changed for the different processes that are working concurrently in a
parallel computing environment.

\section{OpenMP}
\label{sec:openmp}
\texttt{OpenMP} represents the standard API for shared memory parallel
processing. An overview of the various \texttt{OpenMP} functions is
provided, e.g., in \citet{chapman}. Here we will use only the API
features that allow us to parallelize a block of code.

The listing below illustrates the use of \texttt{RngStreams} in an
\texttt{OpenMP} program. The runtime-sized array in this and later
C++ listings uses a GNU compiler extension. These programs also
assume that the runtime supplies the requested number of threads.
\lstset{language=C++,basicstyle={\ttfamily\footnotesize},
  showstringspaces=false,numbers=none}
\begin{lstlisting}[frame=lines,caption={ranOpenMP.cpp}, label=ranopenmp]
//ranOpenMP.cpp
#include <omp.h>
#include "RngStream.h"
#include <iostream>

int main(){

  int nP = omp_get_num_procs();
  omp_set_num_threads(nP);//set number of threads

  unsigned long seed[6] ={1806547166, 3311292359,
			  643431772, 1162448557,
			  3335719306, 4161054083};
  RngStream::SetPackageSeed (seed);
  RngStream RngArray[nP];//array of RngStream objects

  int myRank;
#pragma omp parallel private(myRank)
  {
    myRank = omp_get_thread_num();
#pragma omp critical
    {
      std::cout << RngArray[myRank].RandU01() << " ";
    }
  }
  std::cout << std::endl;
  return 0;
}
\end{lstlisting}
The first step is to include the \texttt{OpenMP} header file
\texttt{omp.h} via an include directive. The number of processors
\texttt{nP} is determined and used to set the number of threads (with
\texttt{omp\_get\_num\_procs} and \texttt{omp\_set\_num\_threads},
respectively).  The \texttt{SetPackageSeed} function is then used to
set the seeds for the \texttt{RngStreams} package. This particular
choice of seeds was made to provide results that can be compared to
those obtained from \texttt{R} in Section \ref{sec:r}.

The desired random deviates are produced with an array of
\texttt{RngStream} objects: one object for each thread.  The
\texttt{RngStream} class default constructor is designed so that the
first object will have the seed designated with
\texttt{SetPackageSeed} while subsequent objects will have initial
states/seeds have been advanced $2^{127}$ states from that of their
predecessor.

An alternative approach that one might try here is to simply create a
unique \texttt{RngStream} object for each thread with a
\texttt{private} specification. In such instances the default
constructor will also be called. However, we have experienced cases
where a race condition can lead to threads with duplicate seeds. The
array approach ensures that all objects are created correctly while
effectively exploiting the shared memory feature that allows all
threads access to the array elements.

A random deviate is generated from each of the \texttt{RngStream}
objects using the class method \texttt{RandU01} that invokes the
backbone generator. These values will serve as subsequent comparison
benchmarks. The actual generation of random numbers in the parallel
region is quite simple with each thread using the object in
\texttt{RngArray} that corresponds to its unique thread number
assigned by the \texttt{OpenMP} API. The resulting values are printed
out from within a \texttt{critical} region to avoid garbled output.
We compile and execute Listing \ref{ranopenmp} using the gnu compiler
suite on a four processor machine in the following manner.
\lstset{language=csh,basicstyle=\ttfamily\footnotesize}
\begin{lstlisting}
$ g++ -fopenmp ranOpenMP.cpp RngStream.cpp -o ranOpenMP
$ ./ranOpenMP
0.341106 0.312399 0.166374 0.149433
\end{lstlisting}

\section{MPI}
\label{sec:mpi}
The \texttt{Message Passing Interface} or \texttt{MPI} provides a
number of functions that can be used for inter-processor
communication; see, e.g., \citet{quinn}. Our development here will use
only the simple communication paradigm where a master and worker
processes transmit and receive messages via the \texttt{MPI}
\texttt{send} and \texttt{recv} functions.

As noted in Section~\ref{sec;rngstreams}, the seeds/states of
\texttt{RngStream} objects are advanced by multiplication involving
known matrices that are available as part of the package. A function
that can be used to carry out these multiplications is
\texttt{MatVecModM} that has prototype
\lstset{language=C++,basicstyle={\ttfamily\footnotesize},
showstringspaces=false,numbers=none}
\begin{lstlisting}
void MatVecModM(const double A[3][3], const double s[3],
		double v[3], double m)
\end{lstlisting}
This function will perform multiplication of a $3 \times 3$ array
\texttt{A} times a three-element array \texttt{s} modulo \texttt{m}
and return the result in the three-element array \texttt{v}. For our
purposes we will take \texttt{m} to be either $m_1 = 4294967087$ or
$m_2 = 4294944443$, $A$ to be either $A_1^{2^{127}}$ or $A_2^{2^{127}}$ and
\texttt{s} to be a seed vector that was allocated to a previous
process.  The moduli and matrices (as well as \texttt{MatVecModM}) are
contained in an anonymous namespace within \texttt{RngStream.cpp} under
the names \texttt{m1}, \texttt{m2}, \texttt{A1p127} and
\texttt{A2p127}. Thus, the following function will manually advance
the input array \texttt{seedIn} $2^{127}$ states to \texttt{seedOut}
which occupies the same relative location in the next stream.
\lstset{language=C++,basicstyle={\ttfamily\footnotesize},
  showstringspaces=false,numbers=none}
\begin{lstlisting}
static void AdvanceSeed(unsigned long seedIn[6],
		   unsigned long seedOut[6]){
  double tempIn[6]; double tempOut[6];
  for(int i = 0; i < 6; i++) tempIn[i] = seedIn[i];
  MatVecModM (A1p127, tempIn, tempOut, m1);
  MatVecModM (A2p127, &tempIn[3], &tempOut[3], m2);
  for(int i = 0; i < 6; i++)
    seedOut[i] = tempOut[i];
}
\end{lstlisting}
We wrapped this function with the necessary supporting code from
\texttt{RngStream.cpp}  in a supplemental class \texttt{RngStreamSupp}
whose header file \texttt{RngStreamSupp.h} is included in the code below.

Our \texttt{MPI} implementation of the \texttt{RngStreams} methodology
now takes the following form.
\lstset{language=C++,basicstyle={\ttfamily\footnotesize},
  showstringspaces=false,numbers=none}
\begin{lstlisting}[frame=lines,caption={ranMPI.cpp}, label=ranmpi]
//ranMPI.cpp
#include <iostream>
#include "mpi.h"
#include "RngStream.h"
#include "RngStreamSupp.h"

int main(){
  unsigned long seed[6] ={1806547166, 3311292359,
			  643431772, 1162448557,
			  3335719306, 4161054083};
  MPI::Init();//start MPI
  int myRank = MPI::COMM_WORLD.Get_rank();//get process rank
  int nP = MPI::COMM_WORLD.Get_size();//get number of processes
  if(myRank == 0){
    //generate deviate for master process
    RngStream::SetPackageSeed(seed);
    RngStream Rng;
    double U = Rng.RandU01();
    //now send seeds to the other processes
    unsigned long tempSeed[6];
    for(int i = 1; i < nP; i++){
      RngStreamSupp::AdvanceSeed(seed, tempSeed);
      MPI::COMM_WORLD.Send(tempSeed, 6, MPI::UNSIGNED_LONG,
			   i, 0);
      for(int j = 0; j < 6; j++)
        seed[j] = tempSeed[j];
    }
    std::cout << U << " ";
    //collect the other random deviates
    for(int i = 1; i < nP; i++){
      MPI::COMM_WORLD.Recv(&U, 1, MPI::DOUBLE, i, 0);
      std::cout << U << " ";
    }
    std::cout << std::endl;
 }
  else{
    unsigned long mySeed[6];
    //receive your seed if you are not the master process
    MPI::COMM_WORLD.Recv(mySeed, 6, MPI::UNSIGNED_LONG, 0, 0);
    RngStream::SetPackageSeed(mySeed);
    RngStream Rng;
    double U = Rng.RandU01();
    //send deviate to master process
    MPI::COMM_WORLD.Send(&U, 1, MPI::DOUBLE, 0, 0);
  }
  MPI::Finalize();//end MPI
  return 0;
}
\end{lstlisting}
The idea behind Listing \ref{ranmpi} is straightforward. The master
process uses an initial seed in conjunction with \texttt{AdvanceSeed}
to determine seeds that will produce ``independent'' random number
streams for the other processes. These seeds are communicated to the
processes using the \texttt{send} and \texttt{recv} functions. The
processes then use them to initialize their particular
\texttt{RngStream} objects and generate a random uniform value. The
communication process is then reversed and the worker processes send
their random numbers back to the master for printing.

A typical compile and run sequence for our \texttt{MPI} program might
look like
\lstset{language=csh,basicstyle={\ttfamily\footnotesize}}
\begin{lstlisting}
$ mpicxx ranMPI.cpp RngStream.cpp RngStreamSupp.cpp -o ranMPI
$ mpiexec -np 4 ranMPI
0.166374 0.341106 0.312399 0.149433
\end{lstlisting}
This output agrees with our prior results from the \texttt{OpenMP}
program, up to the order in which threads entered the critical region.

An alternative approach that leads to the same result as in
\texttt{ranMPI.cpp} is to mimic the array formulation that was used in
\texttt{ranOpenMP.cpp}. This reduces the amount of communication
with the expense of more memory usage for the individual processes. We
illustrate this approach in Section \ref{sec:ltm}.

\section{RngStreams in R}
\label{sec:r}
The \texttt{RngStreams} backbone generator is now one of the random
number generator options in \texttt{R}. For example,
\lstset{language=R, basicstyle={\ttfamily\footnotesize}}
\begin{lstlisting}
> RNGkind("L'Ecuyer-CMRG")
> set.seed(123)
> runif(1)
[1] 0.1663742
\end{lstlisting}
sets the \texttt{R} uniform random number generator as MRG32k3a, sets
the session's seed and reproduces one of the random deviates from the
previous two sections.

At this point it is perhaps worthwhile to mention how we have aligned
the seeds that are being used in our \texttt{R} and \texttt{C++}
code. The \texttt{set.seed} function in \texttt{R} uses a single
integer to set the seed for the current uniform random number
generator. The result can then be accessed through the integer vector
\texttt{.Random.seed}. In terms of our example with the MRG32k3a
generator, the vector has seven elements: the first one indicates the
type of generator and the remaining six are the seeds. Thus,
\lstset{language=R, basicstyle={\ttfamily\footnotesize}}
\begin{lstlisting}
> set.seed(123)
> cat(.Random.seed[2:7], "\n")
1806547166 -983674937 643431772 1162448557 -959247990 -133913213
\end{lstlisting}
reveals \texttt{R}'s representation for the six seeds that are
required for MRG32k3a. Initially the presence of negative integers
seems a bit puzzling; the \texttt{RngStreams} seeds are stored as
\texttt{unsigned long} in the code for the package. However, in
\texttt{R} they are treated as 32-bit unsigned integers internally but
otherwise viewed as signed integers with a two's-complement
representation. Thus, one must add $2^{32}$ to any negative integers
that appear in \texttt{.Random.seed} to see the actual seeds that were
used with the generator. For example,
\lstset{language=R, basicstyle={\ttfamily\footnotesize}}
\begin{lstlisting}
> .Random.seed[3] + 2^{32}
[1] 3311292359
\end{lstlisting}
gives us the second seed that appeared in our \texttt{C++}
code. Although this may appear to be an esoteric detail, it is
essential for cross-language output comparison.

Another way to access the MRG32k3a generator is through the
\texttt{rstream} package that allows us to create instances of a
derived class \texttt{rstream.mrg32k3a} for the backbone generator.
By default, subsequent \texttt{rstream.mrg32k3a} objects are
initialized to the next stream. In fact, \texttt{rstream} will return
an error if more than one generator requests a specific seed in the
same R session. To override this behavior one must specify the option
\texttt{force.seed = TRUE}. The following code initializes four
generators and uses them with the \texttt{rstream.sample} function to
each produce a uniform random deviate.
\lstset{language=R, basicstyle={\ttfamily\footnotesize}}
\begin{lstlisting}
> library(rstream)
> rngList <- c(new("rstream.mrg32k3a", seed = c(1806547166,
+                  3311292359, 643431772, 1162448557, 3335719306,
+                  4161054083), force.seed = TRUE),
+              replicate(3, new("rstream.mrg32k3a")))
> sapply(rngList, rstream.sample)
[1] 0.1663742 0.3411064 0.3123993 0.1494334
\end{lstlisting}
Note that we have again produced the four separate streams that
arose in the previous two sections.

The MRG32k3a generator is a built in fixture of the \texttt{parallel}
package that offers essentially the same parallel computing
functionality as can be obtained separately from the \texttt{snow} \citep{snow} and
\texttt{multicore} \citep{multicore} packages. In addition, \texttt{parallel} is now
bundled within the \texttt{R} base distribution which suggests it will
be one of the key ingredients in the evolution of \texttt{R}'s future
parallel processing capabilities.

The function \texttt{makeCluster} from the \texttt{parallel} package
will create a (default, socket) cluster whose seed can then be set
with \texttt{clusterSetRNGStream}. The code below illustrates this process
using a cluster of size four.
\lstset{language=R, basicstyle={\ttfamily\footnotesize}}
\begin{lstlisting}
> library(parallel)
> cl <- makeCluster(4)
> clusterSetRNGStream(cl, 123)
> parSapply(cl, rep(1, 4), runif)
[1] 0.1663742 0.3411064 0.3123993 0.1494334
> stopCluster(cl)
\end{lstlisting}
The random numbers were generated using \texttt{parSapply} that gives
a parallel version of \texttt{sapply}; there are parallel versions of
\texttt{lapply} and \texttt{apply} as well as several other related
options. Upon completion of our task, the cluster is terminated with
\texttt{stopCluster}. The output confirms that we are using the
MRG32k3a generator and that the seeds are being advanced correctly for
each node in the cluster.

The other way that \texttt{parallel} provides multithreading ability
(for non-Windows machines) is through \texttt{mclapply}, which
furnishes a parallelized version of \texttt{lapply} suited for a
shared memory environment. An attempt to replicate our cluster results
using this approach produces
\lstset{language=R, basicstyle={\ttfamily\footnotesize}}
\begin{lstlisting}
> library(parallel)
> RNGkind("L'Ecuyer-CMRG")
> set.seed(123)
> unlist(mclapply(rep(1,3),runif))
[1] 0.3411064 0.3123993 0.9712727
> runif(1)
[1] 0.1663742
\end{lstlisting}
With the default of two prescheduled child processes, the first
and third tasks run in the same child and use successive values from
its stream. Thus 0.9712727 is the second draw from that child's
stream, rather than the first draw from a third stream. The call
\texttt{mclapply(rep(1,3),runif)} makes three calls to
\texttt{runif} with \texttt{n=1} and collects the results in a list.
To assign these three tasks to three child processes, use
\lstset{language=R, basicstyle={\ttfamily\footnotesize}}
\begin{lstlisting}[frame=lines,caption=mclapply with built-in L'Ecuyer CMRG,label=q:parallel]
> library(parallel)
> RNGkind("L'Ecuyer-CMRG")
> set.seed(123)
> unlist(mclapply(rep(1, 3), runif, mc.cores = 3))
[1] 0.3411064 0.3123993 0.1494334
> runif(1)
[1] 0.1663742
\end{lstlisting}
The \texttt{mc.cores} option sets the maximum number of child
processes used concurrently. This
value is not automatically set to the length of the first argument of
\texttt{mclapply}. On Windows platforms, \texttt{mc.cores} must be set to 1 since \texttt{mclapply} relies on the Unix fork command, which is not available in Windows.

It is possible to gain more direct control over the behavior of the
random streams on each core using the \texttt{rstream} package. This
provides a valuable option when reproducibility across multiple
languages and platforms is desired. For example,
\lstset{language=R, basicstyle={\ttfamily\footnotesize}}
\begin{lstlisting}[frame=lines,caption=mclapply with rstream,label=q:rstream]
> library(rstream)
> library(parallel)
> rngList <- c(new("rstream.mrg32k3a",
+                  seed = c(1806547166, 3311292359, 643431772,
+                           1162448557, 3335719306, 4161054083),
+                  force.seed = TRUE),
+              replicate(3, new("rstream.mrg32k3a")))
> unlist(mclapply(rngList, rstream.sample, mc.cores = 4))
[1] 0.1663742 0.3411064 0.3123993 0.1494334
\end{lstlisting}
returns the results we had anticipated. On Windows platforms, where
\texttt{mc.cores} must be 1, the program in Listing
\ref{q:parallel} will produce numbers from a single stream from
\texttt{RngStreams}. By contrast, the code from Listing
\ref{q:rstream} will faithfully reproduce the results from a multicore
non-Windows platform during serial execution on a Windows machine.
A similar approach of explicitly creating the streams with
\texttt{rstream} may also be used with the cluster approach in order
to maximize transparency.

\section{An application}
\label{sec:ltm}
In this section the basic ideas in Sections
\ref{sec:openmp}--\ref{sec:r} are applied to a real problem involving
Monte Carlo integration. The resulting code listings now become rather
lengthy with the consequence that we can only discuss a few key
aspects here.

The setting is that of \citet{bandl} where a latent trait model (LTM)
is being fit to data obtained from subject responses on the
LSAT, Section VII. There are five dichotomous variables $Y = (Y_1, \ldots, Y_5)$
whose observed values $y = (y_1, \ldots, y_5)$ produce $2^5 = 32$
possible response patterns. If $y$ is one such pattern, it is assumed
that the conditional probability of seeing $y$ given the value $z$ of
a latent variable $Z\sim N(0,1)$ takes the form
\begin{equation}
\label{eq:model1}
p(y|z) = \prod_{i = 1}^5 p_i(z)^{y_i}(1 - p_i(z))^{1 - y_i},
\end{equation}
where
\begin{equation}
\label{eq:model2}
p_i(z) = 1/\left(1 + \exp\left\{\alpha_i - \beta z\right\}\right)
\end{equation}
with $\alpha = (\alpha_1, \ldots, \alpha_5)$ a vector of variable
specific intercepts and $\beta$ a common slope parameter. The sign
convention in (\ref{eq:model2}) agrees with the code below. Reversing
the sign of the common slope leaves the marginal probabilities
unchanged because the standard normal density is symmetric. This leads
to
\begin{eqnarray}
\label{eq:model3}
p(y) &:=& \text{E}_z[p(y|z)] \nonumber\\
&=& \int_{-\infty}^{\infty} p(y|z)\phi(z)dz
\end{eqnarray}
with $\phi$ the standard normal density. We will focus on
approximation of this integral via Monte Carlo methods. Other
integrals such as those obtained by differentiation of
(\ref{eq:model3}) with respect to $\alpha$ and $\beta$ that arise in
the computation of marginal maximum likelihood estimators can be
handled analogously.

The choice of $\alpha, \beta$ is arbitrary for our purposes and,
accordingly, we use the marginal likelihood estimates from a
common-slope fit to the Section VII data, obtained with the
\texttt{grm} function in the \texttt{R} \texttt{ltm} package
\citep{ltm} using 32-point Gauss--Hermite quadrature. The values
shown in the code are the raw thresholds and common slope,
corresponding to \texttt{constrained = TRUE} and
\texttt{IRT.param = FALSE}. What requires somewhat more justification is our focus on an
example with a single latent variable. In general, there may be many
latent variables in an LTM that make the integral in (\ref{eq:model3})
multidimensional. As detailed in Section 1.1 of \citet{lemieux}, the
number of points in product-rule quadrature must grow exponentially
with the dimension of the integrand to maintain a constant rate of
decay for the approximation error. For independent sampling and an integrand with finite variance,
the root mean squared error of a Monte Carlo sample mean is
$\sigma/\sqrt{n}$, where $n$ is the number of sampling points and
$\sigma^2$ is the variance of one importance-weighted evaluation
$\phi(z)p(y|z)/q(z)$ when sampling from a density $q$.
The exponent $-1/2$ does not depend explicitly on dimension, although
$\sigma$ and the cost of an evaluation may do so. Thus, one concludes that Monte Carlo integration is more
appropriate in higher dimensions and, for example, univariate
integrals are better and more precisely evaluated by deterministic
quadrature. On the other hand, the basic premise that underlies Monte
Carlo integration remains the same regardless of the dimension:
repeatedly generate values from a probability distribution, evaluate
the integrand at these values and average the results in some general
sense. Consequently, there is little conceptual downside to our use of
a single latent variable in this instance while the payoff in terms of
code length and mathematical simplification is rather substantial.

For any specified pattern $y$, a Monte Carlo estimator of
(\ref{eq:model3}) will typically have the form
\begin{equation}
\label{eq:probfunc}
\mathrm{probFunc}(y) = \sum_{i = 1}^nW(z_i)p(y|z_i)
\end{equation}
where the sampling distribution and weight must be chosen together.
In the implementation below, $z_1,\ldots,z_n$ are sampled from
$N(0,4)$, with density $q(z)=\phi(z/2)/2$, and
\[ W(z)=\frac{\phi(z)}{nq(z)}=\frac{2}{n}\exp(-3z^2/8). \]
Thus (\ref{eq:probfunc}) is an importance-sampling estimator of
(\ref{eq:model3}). The serial aspect of our computation problem is
primarily about evaluation of (\ref{eq:probfunc}). We will therefore
deal with that issue first before turning to the parallelization step.
In both cases the \texttt{R} and \texttt{C++} code will be developed
in tandem to better illustrate the similarities and differences that
arise when attempting to solve this type of problem in the two
different languages.

The \texttt{C++} code (included in \texttt{funclass.cpp} in the
supplementary material) that evaluates (\ref{eq:probfunc}) is
\lstset{language=C++,basicstyle={\ttfamily\footnotesize},showstringspaces=false}
\begin{lstlisting}[frame=lines,caption={probFunc in C++}, label=probfuncc]
TNT::Array1D<double>
funClass::probFunc(const TNT::Array1D<double>& Z,
		   const TNT::Array2D<int>& patn){
  //pattern probability array used in intermediate calculations
  TNT::Array2D<double>  probMat(patn.dim1(), patn.dim2(), 0.);
  //work array
  TNT::Array1D<double> temp(patn.dim1(), 0.);
  //array that will hold the cell probabilities
  TNT::Array1D<double> prob(patn.dim1(), 0.);

  //sum across normal deviates
  for(int k = 0; k < Z.dim(); k++){
    probMat = probMatFunc(Z[k], patn);

    //multiply to get the probability for each pattern
    for(int i = 0; i < patn.dim1(); i++){
      temp[i] = 1.;
      for(int j = 0; j < patn.dim2(); j++)
	temp[i] *= probMat[i][j];
      temp[i] *= 2.*dnorm(Z[k])
	/(dnorm(Z[k]/2.)*(double) Z.dim());
    }
    prob += temp;
  }
  return prob;
}
\end{lstlisting}
This function takes and returns arguments from the Template Numerical
Toolkit (TNT) that provides implementations of vector (as
\texttt{Array1D} objects) and matrix (as \texttt{Array2D} objects)
classes with overloaded addition, multiplication, etc.,
operators. These array classes are implemented as templates which has
the consequence that all code is placed in header files that can be
downloaded directly from \url{http://math.nist.gov/tnt}.

The call to \texttt{probMatFunc} that appears here corresponds to a
function that evaluates $p_i(z)^{y_i}(1 - p_i(z))^{1 - y_i}$ across
all (32 in this instance) possible response patterns for a given value
of $z$ and returns the result as an array with columns representing
the different variables in the model. The values that are used for the
$z$ argument are passed into the function in the \texttt{Array1D}
object \texttt{Z}. These are chosen to be pseudo-random normal
deviates with 0 mean and standard deviation 2. This feature in
conjunction with the weights that are applied at the end of each pass
through the \texttt{i} loop produces a simple importance sampler
estimate of the target integral. The code for \texttt{probFunc},
\texttt{probMatFunc} and several utility functions (e.g., the standard
normal density and quantile function shown as \texttt{dnorm} and
\texttt{qnorm} in Listing~\ref{probfuncc}) have been collected into a
\texttt{C++} class \texttt{funClass} that contains a class member to
hold the values of the coefficients. In the supplementary
\texttt{funclass.cpp}, the density function is named \texttt{pnorm};
this is a normal density, unlike the distribution function of that
name in \texttt{R}.

The \texttt{R} version of Listing \ref{probfuncc} (see
\texttt{LTM\_parallel.R} in the supplement) is embodied in three
functions. First, there is
\lstset{language=R, basicstyle={\ttfamily\footnotesize}}
\begin{lstlisting}
> PFunc <- function(z, coefVec, patn) {
+   ncells <- nrow(patn)
+   probMat <- ProbMatFunc(z, coefVec, patn)
+   p <- matrix(0, ncells, 1)
+   # multiply the probabilities across variables
+   p <- as.matrix(dnorm(z) * apply(probMat, 1, prod), ncells, 1)
+   p
+ }
\end{lstlisting}
that evaluates the integrand $p(y|z)\phi(z)$ of
(\ref{eq:model3}) across all response patterns (i.e.,
$y$ values) for a given value of the latent variable (i.e., $z$). The
call to \texttt{ProbMatFunc} in this instance is for the \texttt{R}
analog of its \texttt{C++} namesake in Listing \ref{probfuncc}. The
\texttt{patn} argument is an input array that holds the response
patterns as in the \texttt{C++} case. However, in contrast to the
\texttt{C++} development, in this instance the coefficient vector that
is needed for evaluating (\ref{eq:model2}) is treated as an input
variable, named \texttt{coefVec}, that is passed to this and other
functions that use it directly from a driver program.

Next, the function \texttt{PFunc} is vectorized via
\lstset{language=R, basicstyle={\ttfamily\footnotesize}}
\begin{lstlisting}
> VecPFunc <- Vectorize(PFunc, vectorize.args = "z")
\end{lstlisting}
to handle the array case and thereby allow us to use it with
\texttt{apply} for ``averaging'' across normal deviates as in

\lstset{language=R,basicstyle={\ttfamily\footnotesize},showstringspaces=false}
\begin{lstlisting}[frame=lines,caption={ProbFunc in R}, label=probfuncr]
> ProbFuncparLapply <- function(nSim, nP, coefVec, patn) {
+   Z <- qnorm(runif(round(nSim/nP, 0)), 0, 2)
+   prob <- colMeans(apply(VecPFunc(Z, coefVec = coefVec,
+                                   patn = patn),
+                          1, function(y) y/dnorm(Z, 0, 2)))
+ }
\end{lstlisting}
The \texttt{Z} array that appears in this listing is, again, a
vector of pseudo-random normal deviates with 0 mean and standard
deviation 2 that is used in the same weighting scheme as for the
\texttt{C++} code.

The \texttt{C++} \texttt{MPI} code (available from
\texttt{driverLTM\_mpi.cpp} in the supplement) that calculates
(\ref{eq:probfunc}) in parallel looks like
\lstset{language=C++,basicstyle={\ttfamily\footnotesize},
  showstringspaces=false}
\begin{lstlisting}
  RngStream::SetPackageSeed(seed);
  RngStream RngArray[nP];

  //array for random numbers
  TNT::Array1D<double> Z(nSim[myRank], 0.);
  //funClass object
  funClass f(pAlpha, beta, patn.dim2());

  //array that will hold approximate probabilities
  TNT::Array1D<double> prob(patn.dim1(), 0.);

  //approximate the cell probabilities on each process
  for(int i = 0;i < nSim[myRank]; i++)
    Z[i] =  2.*f.qnorm(RngArray[myRank].RandU01());
  prob = f.probFunc(Z, patn);

  //now gather up the results on the master process
  double* temp;
  temp = new double[nP*patn.dim1()];

  double* pData_ = new double[patn.dim1()];
  for(int i = 0; i < patn.dim1(); i++)
    pData_[i] = prob[i];

  MPI::COMM_WORLD.Gather(pData_, patn.dim1(), MPI::DOUBLE,
			 temp, patn.dim1(), MPI::DOUBLE, 0);
  if(myRank == 0){
    std::cout << "Integration results:"<<std::endl;
    for(int i = 0; i < prob.dim(); i++){
      for(int j = 1; j < nP; j++)
         prob[i] += temp[j*prob.dim() + i];
      prob[i] /= (double)nP;
      std::cout<<prob[i]<<std::endl;
    }
    std::cout << "time with " << nP << " processes was " <<
      MPI::Wtime() - wtime << std::endl;
  }
\end{lstlisting}
Here we have employed the scheme from Section \ref{sec:openmp} for
obtaining separate streams: each process creates the same array
of \texttt{RngStream} objects and then uses the array element that
corresponds to its rank. The number of normal deviates that must be
generated by the different processes is portioned out in an array
\texttt{nSim} that is used in the same manner. The \texttt{funClass}
object \texttt{f} then provides the means for each process to
transform the uniform deviates obtained from the \texttt{RngStream}
object into normal deviates with mean 0 and standard deviation 2,
and call \texttt{probFunc} to approximate
the integral.

The use of a common \texttt{RngStream} object array has allowed us to
circumvent the need for sending information to the processes. It is
only necessary to collect up the results they produce. One means to
accomplish this directly, without looping, is by use of \texttt{MPI}'s
\texttt{Gather} function which is the path we have chosen in this
instance. All processes call this function with a pointer argument for
the quantity that is to be passed to the master (or 0) process. This
data is then received (by the master) in a pointer of dimension
sufficient to hold the collective result (i.e., the product of the
number of processes \texttt{nP} and the length \texttt{prob.dim()} of
the \texttt{prob} array). The pointer that is passed from the worker
processes needs to be one that points to the actual data that is held
in the \texttt{prob} array. This resides in a pointer named
\texttt{data\_} that is a private member of the \texttt{Array1D}
class. However, its content is accessible through an overloaded
\texttt{[]} operator and we used this feature to manually transfer the
information in \texttt{data\_} to the pointer \texttt{pData\_} that
was passed to \texttt{Gather}.  Once all the integral approximations
are collected, the master process averages them across processes to
obtain the final approximation.

The \texttt{OpenMP} analog of our \texttt{MPI} code is shown in the
next listing, and is available from \texttt{driverLTM\_openmp.cpp} in
the supplement.
\lstset{language=C++,basicstyle={\ttfamily\footnotesize},
  showstringspaces=false}
\begin{lstlisting}
  //pointer to Array1D objects that will hold output
	//from each process
  TNT::Array1D<double>* tempVec = new TNT::Array1D<double>[nP];

  //pointer for random numbers and funClass object
  double* pZ;
  funClass* pf;
  //initialize array for cell probabilities
  TNT::Array1D<double> prob(patn.dim1(), 0.);

 //approximate the cell probabilities
#pragma omp parallel private(myRank, pZ, pf)
  {
    myRank = omp_get_thread_num();
    pZ = new double[nSim[myRank]];
    pf = new funClass(pAlpha, beta, patn.dim2());
    for(int i = 0;i < nSim[myRank]; i++)
      pZ[i] =  2.*pf->qnorm(RngArray[myRank].RandU01());
    tempVec[myRank]
      = pf->probFunc(TNT::Array1D<double>(nSim[myRank],
					  pZ), patn);
    delete[] pZ;
    delete pf;
  }

  for(int i = 0; i < nP; i++){
    prob += tempVec[i];
  }
  std::cout << "Integration results:"<<std::endl;
  std::cout << std::fixed;
  for(int i = 0; i < patn.dim1(); i++){
    prob[i] /= (double)nP;
    std::cout<<std::setprecision(9)<<prob[i]<<std::endl;
  }

  std::cout <<"time with " << nP
    << " processes was " << omp_get_wtime() - wtime
    << std::endl;
\end{lstlisting}
An array of \texttt{RngStream} objects is used to generate the random
uniforms for each process exactly as in Section \ref{sec:openmp}. What
differs from the \texttt{MPI} treatment is the way we have used
pointers for the \texttt{funClass} object, the array of normal deviates,
and storage of the output from \texttt{probFunc}. By designating the
first two of these pointers as \texttt{private} in the
\texttt{private} clause at the start of the parallel section, we avoid
any race conditions that might arise in calling constructors. Each
thread can safely create the object and array they will need to
perform their task. The now familiar mapping of a process's
rank to an array element index is used to store the \texttt{Array1D}
objects that each thread produces with \texttt{probFunc} in a safe
location provided by the \texttt{Array1D} pointer. After all threads
collapse back to the master at the end of the parallel region, their
output is averaged using an overloaded addition assignment operator
for the \texttt{Array1D} class.

In the \texttt{R} setting our parallel implementation takes the form
\lstset{language=R, basicstyle={\ttfamily\footnotesize}}
\begin{lstlisting}
> library(parallel)
> DriverLTMparLapply <- function(patn, coefVec, nSim, nP) {
+   cl <- makeCluster(nP)
+   clusterSetRNGStream(cl, 123)
+   clusterExport(cl, c("patn", "coefVec", "nSim", "nP"))
+   clusterExport(cl, c("ProbMatFunc", "PFunc", "VecPFunc",
+                       "ProbFuncparLapply"))
+   result <- parLapply(cl, seq_len(nP),
+                       function(...) ProbFuncparLapply(nSim,
+                                           nP, coefVec, patn))
+   stopCluster(cl)
+   # now average across threads
+   prob <- vector(mode = "double", length = nrow(patn))
+   for (i in 1:nP) {
+     prob <- prob + result[[i]]
+   }
+   prob/nP
+ }

\end{lstlisting}
in a cluster context and
\lstset{language=R, basicstyle={\ttfamily\footnotesize}}
	\begin{lstlisting}
> library(parallel)
> library(rstream)
> DriverLTMmclapply <- function(patn, coefVec, nSim, nP) {
+   rngList <- c(new("rstream.mrg32k3a",
+                    seed = c(1806547166, 3311292359, 643431772,
+                             1162448557, 3335719306, 4161054083),
+                             force.seed = TRUE),
+                replicate(nP - 1, new("rstream.mrg32k3a")))
+   result <- mclapply(rngList, ProbFuncmclapply, nSim, nP,
+                      coefVec, patn, mc.cores=nP)
+   # now average across processes
+   prob <- vector(mode = "double", length = nrow(patn))
+   for (i in 1:nP) {
+     prob <- prob + result[[i]]
+   }
+   prob/nP
+ }

\end{lstlisting}
with
\lstset{language=R, basicstyle={\ttfamily\footnotesize}}
\begin{lstlisting}
> ProbFuncmclapply <- function(rngObj, nSim, nP, coefVec, patn) {
+   Z <- qnorm(rstream.sample(rngObj, round(nSim/nP, 0)), 0, 2)
+   prob <- colMeans(apply(VecPFunc(Z, coefVec = coefVec,
+                                   patn = patn),
+                          1, function(y) y/dnorm(Z, 0, 2)))
+ }

\end{lstlisting}
for shared memory purposes. In the cluster example, the exported
objects are defined in the global environment by the accompanying
script; a driver used with local objects should specify their
environment with the \texttt{envir} argument to
\texttt{clusterExport}. The various constants and functions must be
communicated to the workers in a cluster with \texttt{clusterExport}.
We have used the \texttt{rstream}
package in the same manner as Section \ref{sec:r} to ensure proper
stream initialization when using \texttt{mclapply}. Both
\texttt{parLapply} and \texttt{mclapply} return lists that hold the
output from each of the workers. The list elements are then averaged to
obtain the final approximation.

The programs produce matching values for the integral approximations,
to the precision they print, when the number of workers is held constant
and \texttt{nSim} is divisible by that number. Otherwise the C++
program distributes the remainder among workers, whereas the R
program rounds each worker's sample count; matching allocations
would then be needed for a direct output comparison. For example, using four threads with the R \texttt{mclapply} function produces
\lstset{language=R, basicstyle={\ttfamily\footnotesize}}
\begin{lstlisting}
> DriverLTMmclapply(patn,coefVec,nSim,4)
 [1] 0.010591032 0.014465339 0.003321068 0.007459930 0.008497401
 [6] 0.019087239 0.004382200 0.016301170 0.004348307 0.009767360
[11] 0.002242468 0.008341667 0.005737658 0.021343283 0.004900161
[16] 0.031011491 0.012769320 0.028683011 0.006585274 0.024496295
[21] 0.016849315 0.062677083 0.014389904 0.091068923 0.008622165
[26] 0.032073241 0.007363630 0.046601970 0.018840844 0.119237442
[31] 0.027375482 0.310245431
\end{lstlisting}
The C++ OpenMP program with four threads produces the same values
(output reformatted five values per line; timing line omitted):
\lstset{language=C++,basicstyle={\ttfamily\footnotesize},
  showstringspaces=false}
\begin{lstlisting}
$ g++ -fopenmp RngStream.cpp funclass.cpp RngStreamSupp.cpp
      driverLTM_openmp.cpp -o openmp_ltm
$ ./openmp_ltm 4
Integration results:
0.010591032 0.014465339 0.003321068 0.007459930 0.008497401
0.019087239 0.004382200 0.016301170 0.004348307 0.009767360
0.002242468 0.008341667 0.005737658 0.021343283 0.004900161
0.031011491 0.012769320 0.028683011 0.006585274 0.024496295
0.016849315 0.062677083 0.014389904 0.091068923 0.008622165
0.032073241 0.007363630 0.046601970 0.018840844 0.119237442
0.027375482 0.310245431
\end{lstlisting}
Though not shown, the R \texttt{parLapply} program produces the same
values, and the C++ MPI program prints them to its default six
significant digits. When the number of threads is changed, the approximations change as well since the random numbers are generated from different streams.

We carried out a few performance comparisons using the \texttt{MPI}
and \texttt{OpenMP} \texttt{C++} code and the \texttt{mclapply} and \texttt{parLapply}
\texttt{R} code for the Monte Carlo integration application. The average speedups
(i.e., the ratios of serial to parallel run times) and run times (in
parentheses) are given in Table \ref{tab:t1}. In all cases we used
five runs with $10^5$ sampling points (i.e., \texttt{nSim = $10^5$}) and
used 1, 2, 4 and 8 processors. While all of the programs perform well, near-linear speedup is achieved with \texttt{mclapply} in R (using the \texttt{rstream} package) and with MPI in C++. As expected, the C++ programs execute at least an order of magnitude faster than the R programs.

\begin{table}[h!!!!]
\begin{center}
\caption{Speedups and (run times, in seconds) for \texttt{C++} and \texttt{R} code}
\label{tab:t1}
\begin{tabular}{ccccc}  \\ \hline
Processors  & mclapply&parLapply   & MPI   & OpenMP
\\
\hline
 1 &1.000 &1.000 &1.000 &1.000 \\
&(77.941)&  (82.541) & (2.254) & (2.199) \\
2&2.070 & 1.950 &2.014 & 1.430 \\
& (37.651)&(42.319) & (1.119) & (1.538)\\
4 &4.117&  3.597 & 4.002 & 3.043 \\
& (18.930)&(22.944) & (0.563) &  (0.723) \\
8 &7.884& 5.549 &7.545& 6.595\\
& (9.886)&(14.874)  & (0.299) & (0.333) \\
 \hline
\end{tabular}
\end{center}
\end{table}
As a rate-based illustration, equating the Monte Carlo order
$n^{-1/2}$ with the $m^{-2}$ error order of an $m$-point trapezoidal
rule in one dimension gives $m\approx18$ when $n=10^5$. A product
trapezoidal grid with
18 points in each of eight dimensions has about $10^{10}$ points.
This comparison illustrates the growth in product-rule cost; actual
accuracy also depends on smoothness, the integration domain, variance
and the constants in the error bounds.

\section{Summary}
\label{sec:summary}
The \texttt{RngStreams} software package provides one freely available
solution for allocating separate random number streams for simulation
experiments that are conducted in a parallel computing
environment. The goal of this paper has been to provide an
introduction to its use in both distributed and shared memory
settings. We have provided minimal working examples along with a more
detailed application to illustrate the potential of
\texttt{RngStreams}.

While \texttt{RngStreams} is easy to use, some care is necessary to
ensure that streams are distributed correctly to different
processes. We have accomplished this by using an array of
\texttt{RngStream} objects where the array index is employed to
uniquely determine which process may access each object. With exclusive access to each object and consumption confined to
its allocated block, this scheme assigns non-overlapping streams in
both \texttt{OpenMP} and \texttt{MPI}. Non-overlap alone is not a
proof of probabilistic independence. In the \texttt{MPI} context, we have described another
option that creates seeds on the master process and then distributes
them to all the worker processes. There are memory savings from this
latter approach that come with the cost of additional inter-processor
communication.

In \texttt{R} we demonstrated the use of \texttt{RngStreams} through
both the \texttt{parallel} and \texttt{rstream} packages. We explored
the built-in functionality for \texttt{RngStreams} in
\texttt{parallel} and demonstrated how \texttt{rstream} may be used
within \texttt{parallel} to improve the portability of the software.

\section*{Acknowledgements} Some of the computations
were carried out on the Saguaro cluster at Arizona State
University. The authors are grateful for comments from the anonymous referees
that lead to improvements in the paper.

\bibliographystyle{asabst}
\bibliography{rngstreamsbib}
\end{document}